\documentclass{elsart}

\usepackage{graphicx}
\usepackage[center,footnotesize,hang]{subfigure}
\usepackage{longtable}
\usepackage{amssymb}
\usepackage{amsmath}
\usepackage{amsfonts}
\usepackage{bbm} 
\usepackage{fleqn} 
\usepackage{indentfirst} 
\usepackage{braket}
\usepackage{mathrsfs}
\usepackage{HEP} 
\def\D{\mathrm{d}} 
\def\SU{\text{SU}}
\def\Tr{\text{Tr}}

\def\<{\left\langle}
\def\>{\right\rangle}

\newcommand{\superfield}[1]{\boldsymbol{#1}}	
\def\chargec{\mathrm{C}}

\begin{document}
\renewcommand{\thesubfigure}{%
(\ifcase\expandafter\arabic{subfigure}\or a\or b\or c\or d\or e\or f\or g\or h\or i\or j\or
k\or l\or m\or n\or o\or p\or q\or r\or s\or t\or u\or v\or w\or x\or
y\or z\or aa\or ab\or ac\or ad\or ae\or af\or ag\or ah\or ai\or aj\or
ak\or a\ell\or am\or an\or ao\or ap\or aq\or ar\or as\or at\or au\or av\or
aw\or ay\or az\or ba\or bb\or bc\or bd\or be\or bf\or bg\or bh\or bi\or
bj\or bk\or b\ell\or bm\or bn\or bo\or bp\or bq\or br\or bs\or bt\or
bu\or bw\or bx\or by\or bz\or ca\or cb\or cc\or cd\or ce\or cf\or cg\or
ch\or ci\or cj\or ck\or cl\or cm\or cn\or co\or cp\or cq\or cr\or
cs\or ct\or cu\or cw\or cx\or cy\or cz\or da\or db\or dc\or dd\or de\or df\or dg\or dh\or di\or dj\or dk\or
d\ell\or dm\or dn\or do\or dp\or dq\or dr\or ds\or dt\or du\or dw\or
dx\or dy\or dz\or ea\or eb\or ec\or ed\or ee\or ef\or eg\or eh\or
ei\or ej\or ek\or el\or em\or en\or eo\or ep\or eq\or er\or es\or
et\or eu\or ew\or ex\or ey\or ez\else\@ctrerr\fi)}
\renewcommand{\subfigtopskip}{8pt}
\renewcommand{\subfigbottomskip}{8pt}
\renewcommand{\subfigcapskip}{6pt}

\begin{frontmatter}

\begin{flushright}
{\small TUM-HEP-443/01}
\end{flushright}
\vspace*{2.cm}
\title{Neutrino Mass Operator Renormalization \\ in  
Two Higgs Doublet Models and the MSSM
}
\author{Stefan Antusch\thanksref{label01}},
\thanks[label01]{E-mail: \texttt{santusch@ph.tum.de}}
\author{Manuel Drees\thanksref{label02}},
\thanks[label02]{E-mail: \texttt{drees@ph.tum.de}}
\author{J\"{o}rn Kersten\thanksref{label04}},
\thanks[label04]{E-mail: \texttt{jkersten@ph.tum.de}}
\author{Manfred Lindner\thanksref{label05}},
\thanks[label05]{E-mail: \texttt{lindner@ph.tum.de}}
\author{Michael Ratz\thanksref{label06}}
\thanks[label06]{E-mail: \texttt{mratz@ph.tum.de}}
\address{Physik-Department T30, 
Technische Universit\"{a}t M\"{u}nchen\\ 
James-Franck-Stra{\ss}e,
85748 Garching, Germany
}

\begin{abstract}
In a recent re-analysis of the Standard Model (SM) \(\beta\)-function
for the effective neutrino mass operator, we found that the previous results were
not entirely correct.
Therefore, we consider the analogous dimension five operators in a class of Two Higgs
Doublet Models (2HDM's) and the Minimal Supersymmetric Standard Model
(MSSM).  Deriving the renormalization group equations for these
effective operators, we confirm the existing result in the case of the
MSSM. Some of our 2HDM results are new, while others differ from earlier
calculations.  This leads to modifications in the renormalization group
evolution of leptonic mixing angles and CP phases in the 2HDM's.
\end{abstract}

\end{frontmatter}

\newpage
\section{Introduction}

The discovery of neutrino masses requires an extension of the Standard
Model (SM). The most promising scenario for giving masses to neutrinos
is the see-saw mechanism \cite{seesaw}, which provides a convincing
explanation for their smallness. It typically introduces heavy,
gauge-singlet neutrinos and thereby gives small Majorana masses to the
SM neutrinos. When the SM is viewed as an effective field theory,
Majorana masses for the neutrinos can be introduced via higher
dimensional operators of SM fields. The lowest dimensional operator of
this kind has dimension 5 and couples two lepton and two Higgs doublets.
It appears e.g.\ in the see-saw mechanism by integrating out the heavy
singlets.

The experimental results in the neutrino sector provide an
interesting new way for testing theories beyond the SM. In order to
compare the experimental results with predictions from models beyond the
SM, like unified theories, it is essential to evolve the masses, mixing
angles and CP phases from high to low energies. This is accomplished
with the renormalization group equations (RGE's) for the neutrino mass
operators in the theory valid at intermediate energy scales. This theory
may be the SM, but can also be an extension like a Two Higgs Doublet
Model (2HDM) or the Minimal Supersymmetric Standard Model (MSSM). 

In a recent letter \cite{Antusch:2001ck} we discussed the derivation of
the RGE for the dimension 5 neutrino mass operator in the SM. In this
letter we derive the RGE's for the corresponding operators in a
class of 2HDM's and the MSSM. 

\section{Effective Neutrino Mass Operators in 2HDM's}

In many extensions of the SM, the Higgs
sector is enlarged by introducing additional 
\(\mathrm{SU}(2)_\mathrm{L}\) doublet scalar fields \(\phi^{(i)}\) 
(\(1\le i\le N_\mathrm{H}\)).
These can couple to the SM fermions via the Yukawa couplings
\begin{eqnarray}
 \mathscr{L}^{(i)}_\mathrm{Yukawa}
 & = &
  -(Y_e^{(i)})_{gf} \overline{e_\mathrm{R}^g}
 	\phi^{(i)\dagger}_a \delta^{ab} \ell_{\mathrm{L}b}^f
 \nonumber\\
 & &
 -(Y_d^{(i)})_{gf}\overline{d_\mathrm{R}^g}
 	\phi^{(i)\dagger}_a \delta^{ab}Q_{\mathrm{L}b}^f
 -(Y_u^{(i)})_{gf}\overline{u_\mathrm{R}^g}
 	Q_{\mathrm{L}b}^f\varepsilon^{ba}\phi^{(i)}_a
 \;.
	\label{eq:THDMYukawaCouplings}
\end{eqnarray}
$\ell_\mathrm{L}^f$ and $Q_\mathrm{L}^f$, $f\in\{1,2,3\}$ are the 
$\SU(2)_\mathrm{L}$ doublets of SM leptons and quarks, respectively.
$e_\mathrm{R}^f$, $d_\mathrm{R}^f$ and $u_\mathrm{R}^f$ denote the
$\SU(2)_\mathrm{L}$-singlet
(right-handed) charged leptons, down-type quarks and up-type quarks.
$\varepsilon$ is the totally antisymmetric tensor in 
2 dimensions and $a,b,c,d \in \{1,2\}$ 
are SU(2) indices. Summation over repeated indices 
is implied throughout this letter.
We have chosen the notation in equation \eqref{eq:THDMYukawaCouplings}
in such a way that all \(\phi^{(i)}\) transform as
\((\boldsymbol{2},\tfrac{1}{2})\) under
\(\mathrm{SU}(2)_\mathrm{L} \otimes \mathrm{U}(1)_\mathrm{Y}\). In
particular, for \(N_\mathrm{H}=1\) we obtain the SM. 

Note that there are tight phenomenological constraints on Yukawa couplings. As
point\-ed out in \cite{Weinberg:1976hu,Glashow:1977nt,Paschos:1977ay}, it is very hard
to construct viable models in which one type of SM fermions \(e\), \(d\)
and \(u\) couples to two or more Higgs bosons, since this in general
leads to tree-level flavor-changing neutral currents (FCNC's).
Therefore, we will only consider models in which the fermions couple to
at most one Higgs. As a consequence, the suffix ``$(i)$'' on the Yukawa
couplings in equation \eqref{eq:THDMYukawaCouplings} is redundant and
will be omitted in the following.

\subsection{Classification of 2HDM's}
We concentrate on models with two Higgs doublets for simplicity, i.e.\
\(N_\mathrm{H}=2\), and consider only schemes in which each of the
right-handed SM fermions couples to exactly one Higgs boson. All
inequivalent possibilities are classified in table
\ref{tab:ClassificationOf2HDM}. By convention, the scalar which couples
to \(e\) is defined to be \(\phi^{(1)}\). 
\begin{table}[h]
\begin{longtable}[c]{lcccc}
\begin{tabular}{@{}l@{}}
Coupling\\[-0.3cm]
scheme
\end{tabular}
&
\(\vcenter{\hbox{\includegraphics{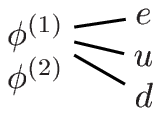}}}\)
&
\(\vcenter{\hbox{\includegraphics{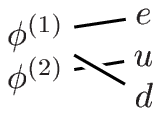}}}\)
&
\(\vcenter{\hbox{\includegraphics{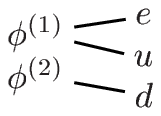}}}\)
&
\(\vcenter{\hbox{\includegraphics{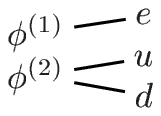}}}\)
\\
Model & (i) & (ii) & (iii) & (iv) \\
\caption{Classification of the 2HDM's with natural suppression of FCNC's and
tree-level mass terms for all SM fermions except neutrinos. Note that model (i) is usually
referred to as ``type I'' and (ii) as ``type II'' in the literature.}
\label{tab:ClassificationOf2HDM}
\end{longtable}
\end{table}

In order to avoid FCNC's, we impose
the \(\mathbbm{Z}_2\) symmetry
\begin{equation}\label{eq:DiscreteSymmetriesOf2HDM}
 \phi^{(1)}\to \phi^{(1)} \;, \qquad
 \phi^{(2)}\to -\phi^{(2)}
\end{equation}
and corresponding transformations in the fermion sector.
For example, in scheme (ii) all fields transform trivially except for
\begin{equation}\label{eq:DiscreteSymmetriesOfTypeii}
 \left(\begin{array}{c}\phi^{(2)} \\ u\end{array}\right)
 \to 
 -\left(\begin{array}{c}\phi^{(2)} \\ u\end{array}\right)
 \;.
\end{equation}

The most general Higgs self-interaction Lagrangian is then
\begin{eqnarray}
 \mathscr{L}_{2\mathrm{Higgs}}
& = &
 -\frac{\lambda_1}{4} \left(\phi^{(1)\dagger} \phi^{(1)}\right)^2
 -\frac{\lambda_2}{4} \left(\phi^{(2)\dagger} \phi^{(2)}\right)^2
\nonumber\\
& &
 -\lambda_3 \left(\phi^{(1)\dagger} \phi^{(1)}\right)
 	\left(\phi^{(2)\dagger} \phi^{(2)}\right) 
 -\lambda_4 \left(\phi^{(1)\dagger} \phi^{(2)}\right)
 	\left(\phi^{(2)\dagger} \phi^{(1)}\right) 
\nonumber\\
& &
 -\Bigl[\frac{\lambda_5}{4} \left(\phi^{(1)\dagger} \phi^{(2)}\right)^2
 +\text{h.c.} \Bigr] \;.
 \label{eq:ScalarPotentialOf2HDM}
\end{eqnarray}

\subsection{Effective Neutrino Mass Operators}

The lowest dimensional effective neutrino mass operators compatible with
the symmetry \eqref{eq:DiscreteSymmetriesOf2HDM} are given by
\begin{equation}
 \mathscr{L}_{\kappa} =
 \mathscr{L}_{\kappa}^{(11)} + \mathscr{L}_{\kappa}^{(22)} \;,
\end{equation}
where
\begin{equation}
 \mathscr{L}_{\kappa}^{(ii)} = 
 \frac{1}{4}\kappa^{(ii)}_{gf}\,\overline{\ell_\mathrm{L}^\mathrm{C}}^g_c
 \varepsilon^{cd}\phi^{(i)}_d\,
 \ell_{\mathrm{L}b}^f \varepsilon^{ba} \phi^{(i)}_a
 + \text{h.c.}
 \qquad(i=1,2) \;. 
\end{equation}
$\ell_\mathrm{L}^\chargec$ is the charge conjugate of the lepton
doublet, and $\kappa^{(ii)}_{gf}$ are symmetric matrices with respect to
the generation indices \(g\) and \(f\).
Note that it is possible that only one of these operators, e.g.\
$\mathscr{L}_{\kappa}^{(22)}$, arises from integrating out heavy degrees
of freedom in a specific model. However, as we shall see, both mix
due to the renormalization group evolution and therefore have to be taken into account simultaneously.

As long as the symmetry \eqref{eq:DiscreteSymmetriesOf2HDM} is valid,
$\mathscr{L}_{\kappa}^{(11)}$ and $\mathscr{L}_{\kappa}^{(22)}$
represent the only possible dimension 5 operators containing two
\(\ell_\mathrm{L}\) fields. If this symmetry was broken, further
couplings would appear in the Higgs interaction Lagrangian
\eqref{eq:ScalarPotentialOf2HDM}.

\subsection{Calculation of the RGE}

We work in the MS renormalization scheme at the 
one-loop level. The wavefunction renormalization constants 
$Z_i = 1 + \delta Z_i$ are defined in the usual way. For the Higgs
fields \(\phi^{(i)}\) we obtain
\begin{eqnarray}
 \delta Z_{\phi^{(i)}}
 & = & 
 -\frac{1}{16\pi^2}
 \left[\delta_{i1}\, 2\Tr(Y_e^\dagger Y_e)
 	+z^{(i)}_u\, 6\Tr(Y_u^\dagger Y_u)
	+z^{(i)}_d\, 6\Tr(Y_d^\dagger Y_d)
	\right.
\nonumber\\
 & & \hphantom{-\frac{1}{16\pi^2}\left[\right.}
 \left.{}
 	+\tfrac{1}{2} (\xi_B-3) \,g_1^2 + \tfrac{3}{2} (\xi_W-3) \,g_2^2
 \right]\frac{1}{\epsilon}\;,
\end{eqnarray}
where $\epsilon := 4-d$ is the deviation from 4 dimensions in
dimensional regularization, and
\(\xi_B\) and \(\xi_W\) are the gauge fixing parameters used in
\(R_\xi\) gauge. $g_1$ and $g_2$ are the U(1)$_{\mathrm{Y}}$ and
SU(2)$_{\mathrm{L}}$ gauge coupling constants, respectively. 
The coefficients \(z_f^{(i)}\) are defined to be 1 if the 
fermion \(f\) couples to the Higgs boson \(\phi^{(i)}\) and 0 otherwise.
For the models classified in table \ref{tab:ClassificationOf2HDM}
they are given by table \ref{tab:ZConstantIn2HDMS}.
\begin{table}[h]
\begin{longtable}[c]{l|cccc}
 	& (i) & (ii) & (iii) & (iv) \\
 	\hline
	\(z^{(1)}_u\) & 1 & 0 & 1 & 0\\
	\(z^{(2)}_u\) & 0 & 1 & 0 & 1\\
	\(z^{(1)}_d\) & 1 & 1 & 0 & 0\\
	\(z^{(2)}_d\) & 0 & 0 & 1 & 1\\
\caption{The coefficients \(z_f^{(i)}\) for the 
Two Higgs Doublet Models classified 
in table \ref{tab:ClassificationOf2HDM}.}
\label{tab:ZConstantIn2HDMS}
\end{longtable}
\end{table}

The wavefunction renormalization for \(\ell_\mathrm{L}\) is given by
\begin{equation}
 \delta Z_{\ell_\mathrm{L}} = 
 -\frac{1}{16\pi^2}
 \left[Y_e^\dagger Y_e
 	+\tfrac{1}{2} \xi_B g_1^2 +\tfrac{3}{2}\xi_W g_2^2
 \right]\frac{1}{\epsilon} \;.
\end{equation}

With the counterterms for the effective vertices defined by
\begin{eqnarray}
 \mathscr{C}_{\kappa}
& = & 
 \frac{1}{4} \delta\kappa^{(11)}_{gf}\,\overline{\ell_\mathrm{L}^\mathrm{C}}^g_c
 \varepsilon^{cd}\phi^{(1)}_d\,
 \ell_{\mathrm{L}b}^f \varepsilon^{ba} \phi^{(1)}_a
\nonumber\\
& & {}+
  \frac{1}{4} \delta\kappa^{(22)}_{gf}\,\overline{\ell_\mathrm{L}^\mathrm{C}}^g_c
 \varepsilon^{cd}\phi^{(2)}_d \,
 \ell_{\mathrm{L}b}^f \varepsilon^{ba}\phi^{(2)}_a
 +\text{h.c.} \;,
\end{eqnarray}
we find for the vertex correction 
\begin{eqnarray}
 \delta\kappa^{(ii)}
 & = &
 -\frac{1}{16\pi^2}
 \left[
 	\delta_{i1}\,2\kappa^{(ii)}(Y_e^\dagger Y_e)
	+\delta_{i1}\,2(Y_e^\dagger Y_e)^T \kappa^{(ii)}
 \right.
\nonumber\\
 & & \hphantom{-\frac{1}{16\pi^2} \left[\right.}
 \left.{}
 	-\lambda_i \kappa^{(ii)}
	-\delta_{i1} \lambda_5^*\kappa^{(22)}
	-\delta_{i2} \lambda_5\kappa^{(11)}
 \right.
\nonumber\\
 & & \hphantom{-\frac{1}{16\pi^2} \left[\right.}
 \left.{}
 	+\left(\xi_B-\tfrac{3}{2}\right)\,g_1^2\kappa^{(ii)}
	+\left(3\xi_W-\tfrac{3}{2}\right)\,g_2^2\kappa^{(ii)}
 \right]
 \frac{1}{\epsilon}\;.
\end{eqnarray}
\begin{figure}[hb]
\subfigure[]
   {\label{subfig:Delta1YeKappaII}$
   	\vcenter{\hbox{\includegraphics[scale=0.8]{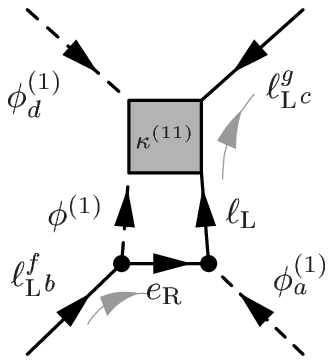}}}
	$}
\hfil
\subfigure[]
   {\label{subfig:Delta2YeKappaII}$
   	\vcenter{\hbox{\includegraphics[scale=0.8]{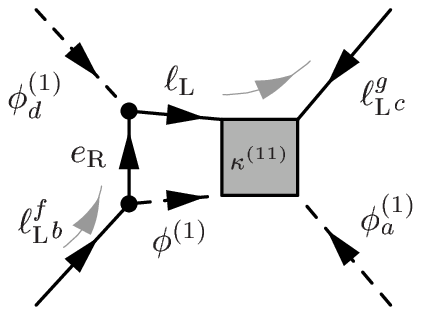}}}
	$}
\hfil
\subfigure[]
   {\label{subfig:Delta3YeKappaII}$
   	\vcenter{\hbox{\includegraphics[scale=0.8]{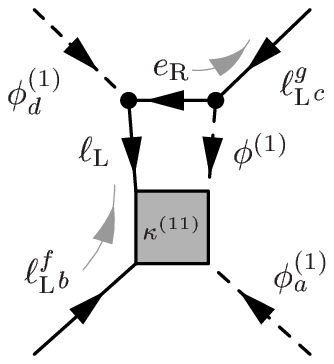}}}
	$}
\hfil
\subfigure[]
   {\label{subfig:Delta4YeKappaII}$
   	\vcenter{\hbox{\includegraphics[scale=0.8]{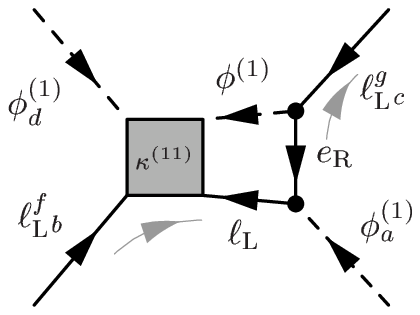}}}
	$}
\\
\subfigure[]
   {\label{subfig:Delta1GaugeKappaII}$
   	\vcenter{\hbox{\includegraphics[scale=0.8]{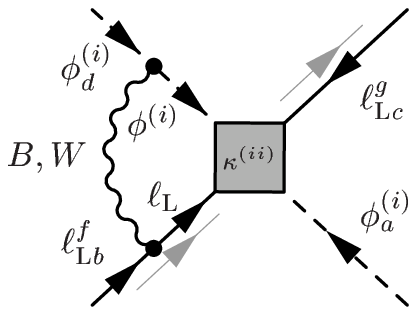}}}
	$}
\hfil
\subfigure[]
   {\label{subfig:Delta2GaugeKappaII}$
   	\vcenter{\hbox{\includegraphics[scale=0.8]{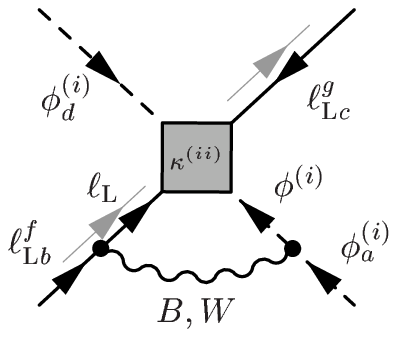}}}
	$}
\hfil
\subfigure[]
   {\label{subfig:Delta3GaugeKappaII}$
   	\vcenter{\hbox{\includegraphics[scale=0.8]{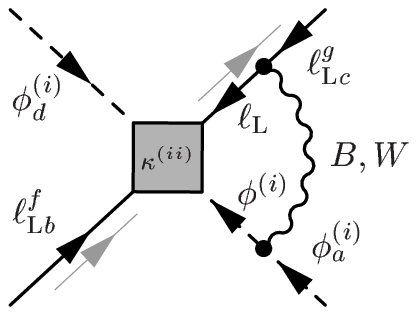}}}
	$}
\\
\subfigure[]
   {\label{subfig:Delta4GaugeKappaII}$
   	\vcenter{\hbox{\includegraphics[scale=0.8]{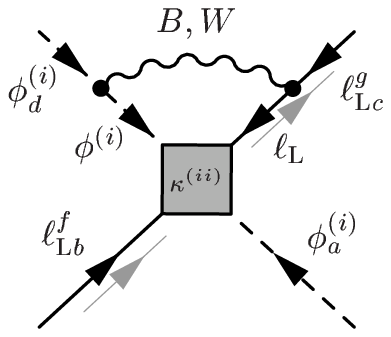}}}
	$}
\hfil
\subfigure[]
   {\label{subfig:Delta5GaugeKappaII}$
   	\vcenter{\hbox{\includegraphics[scale=0.8]{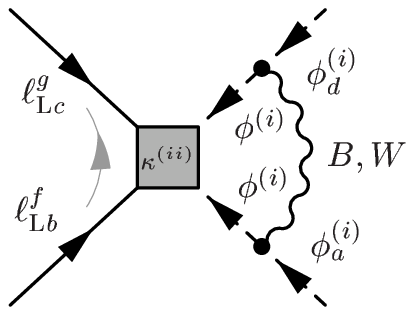}}}
	$}
\hfil
\subfigure[]
   {\label{subfig:Delta6GaugeKappaII}$
   	\vcenter{\hbox{\includegraphics[scale=0.8]{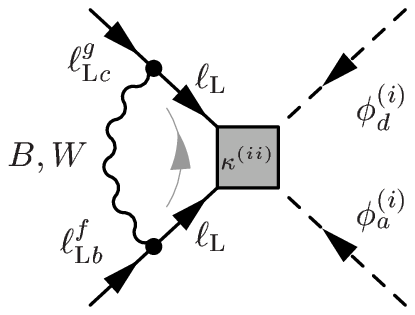}}}
	$}
\\
\subfigure[]
   {\label{subfig:DeltaLambda1Kappa1}
   \(
   	\vcenter{\hbox{\includegraphics[scale=0.8]{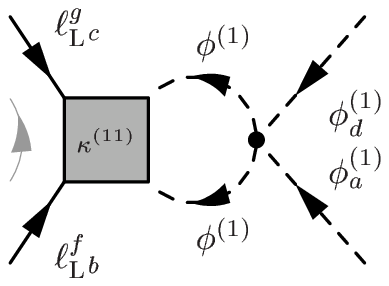}}}
	\)}
\hfil
\subfigure[]
   {\label{subfig:DeltaLambda2Kappa22}
   \(
   	\vcenter{\hbox{\includegraphics[scale=0.8]{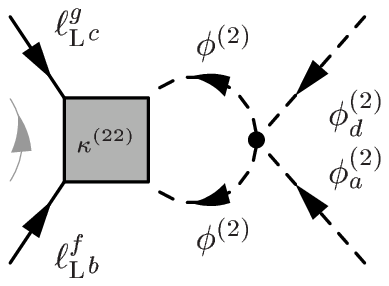}}}
	\)}
\hfil
\subfigure[]
   {\label{subfig:DeltaLambda5Kappa11}$
   	\vcenter{\hbox{\includegraphics[scale=0.8]{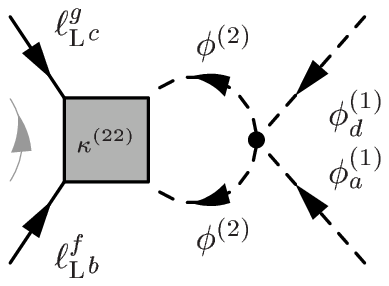}}}
	$}
\hfil
\subfigure[]
   {\label{subfig:DeltaLambda5Kappa22}
   \(
   	\vcenter{\hbox{\includegraphics[scale=0.8]{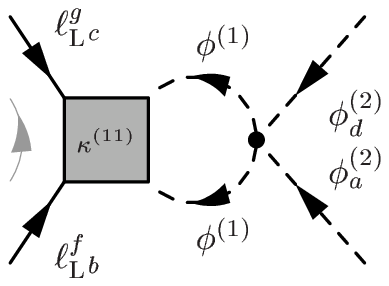}}}
	\)}
\caption{
One-loop diagrams contributing to the vertex renormalization of 
\(\kappa^{(ii)}\).
The one-loop diagrams 
(a) -- (d) that arise due to the Yukawa coupling 
\(Y_e\) affect only the renormalization of \(\kappa^{(11)}\).
The one-loop gauge diagrams 
(e) -- (j) are the same as in the SM. $B$ and 
$W \in \{W^1,W^2,W^3\}$ are the gauge bosons of $\SU(1)_\mathrm{Y}$ and 
$\SU(2)_\mathrm{L}$. Diagrams (k) -- (n)
come from the Higgs interaction Lagrangian.
While the diagrams (k) and (l)
have a counterpart in the SM, the diagrams (m) and (n)
appear only in the 2HDM's and lead to a mixing between 
the operators \(\mathscr{L}_\kappa^{(11)}\) and \(\mathscr{L}_\kappa^{(22)}\).
The gray arrow indicates the fermion flow as defined in \cite{Denner:1992vz}.}
\label{fig:KappaIIVertexCorrections}
\end{figure}

The relevant 
Feynman diagrams are shown in figure \ref{fig:KappaIIVertexCorrections}.
Using the technique described in \cite{Antusch:2001ck}, 
the $\beta$-functions 
$\beta_{\kappa^{(ii)}}=\mu \frac{\D}{\D \mu} \kappa^{(ii)}$
with $\mu$ denoting the renormalization scale,
can be obtained from the counterterms,
\begin{eqnarray}
 16\pi^2\beta_{\kappa^{(ii)}}
 & = & \left(\tfrac{1}{2}-2\delta_{i1}\right)\,
 \left[\kappa^{(ii)}(Y_e^\dagger Y_e)
 	+(Y_e^\dagger Y_e)^T\kappa^{(ii)}\right]
\nonumber\\
 & & {}
 +\left[\delta_{i1}\,2 \,\Tr(Y_e^\dagger Y_e)
 	+z^{(i)}_u\,6\,\Tr(Y_u^\dagger Y_u)
	+z^{(i)}_d\,6\,\Tr(Y_d^\dagger Y_d)\right]\kappa^{(ii)}
\nonumber\\
 & & {}
	+\lambda_i\kappa^{(ii)}+\delta_{i1}\lambda_5^*\kappa^{(22)}
 	+\delta_{i2}\lambda_5\kappa^{(11)}
	-3g_2^2\kappa^{(ii)} \;.
	\label{eq:2HDMBetaFunctions}
\end{eqnarray}
The terms proportional to $\lambda_5$ are responsible for the mixing of
the effective operators mentioned before. 
Our result for $\beta_{\kappa^{(11)}}$ differs from the one in
\cite{Babu:1993qv} by a factor of 3 because of the term
$\tfrac{1}{2}-2\delta_{i1}$ in the first line.
We had earlier found \cite{Antusch:2001ck} an analogous discrepancy 
in \(\beta_\kappa\) for the SM,
which has recently been confirmed \cite{Chankowski:2001hx}.

Note that in 2HDM's running effects are in general
larger than in the SM due to the fact that the Yukawa couplings are 
enhanced, e.g.\ \((Y_e)_{\mathrm{2HDM}}=(Y_e)_{\mathrm{SM}}/\cos\beta\),
where \(\tan\beta=v^{(1)}/v^{(2)}\) with \(v^{(i)}\) being the 
vacuum expectation value of the Higgs field \(\phi^{(i)}\).

\filbreak
\section{The Effective Neutrino Mass Operator in the MSSM}
In the MSSM, the effective dimension 5 operator that gives 
Majorana masses to the SM
neutrinos is contained in the $F$-term
of the superpotential
\begin{eqnarray}
 \mathscr{W}_{\kappa}^{\mathrm{MSSM}} 
 =-\frac{1}{4} 
  {\kappa}^{}_{gf} \, \superfield{L}^{g}_c\varepsilon^{cd}
 \superfield{H}^{(2)}_d\, 
 \, \superfield{L}_{b}^{f}\varepsilon^{ba} \superfield{H}^{(2)}_a 
 +\text{h.c.}\;.
\end{eqnarray}
$\superfield{L}$ and $\superfield{H}^{(2)}$ are 
the chiral superfields that
contain the SU(2)$_\mathrm{L}$ doublets, the Higgs doublet with weak
hypercharge $+\tfrac{1}{2}$ and the corresponding superpartners.
The part of the superpotential describing the 
Yukawa interactions is given by
\begin{eqnarray}
 \mathscr{W}_{\mathrm{Yuk}}^{\mathrm{MSSM}} 
 &=&
 (Y_e)_{gf}\superfield{E}^{\chargec g}
 	\superfield{H}^{(1)}_a\varepsilon^{ab}\superfield{L}^f_b
 \nonumber \\
 && {}
 +(Y_d)_{gf}\superfield{D}^{\chargec g}
 	\superfield{H}^{(1)}_a\varepsilon^{ab}\superfield{Q}^f_b
 +(Y_u)_{gf}\superfield{U}^{\chargec g}
 \superfield{H}^{(2)}_a (\varepsilon^T)^{ab}  \superfield{Q}_b^f \; .	
\end{eqnarray}
The superfields \(\superfield{E}^{\chargec }\), \(\superfield{D}^{\chargec }\)
and \(\superfield{U}^{\chargec}\) contain
the \(\mathrm{SU}(2)_\mathrm{L}\)-singlet charged leptons, 
down-type quarks 
and up-type quarks, respectively, 
and $\superfield{Q}$
contains the SU(2)$_\mathrm{L}$ quark doublets.
The Higgs superfield $\boldsymbol{H}^{(1)}$ has weak
hypercharge $-\tfrac{1}{2}$.
Calculating the RGE in the MSSM yields
\begin{equation}\label{eq:RGEMSSM}
16\pi^2 {\beta}_\kappa^{\mathrm{MSSM}}  =  
 (Y_e^\dagger Y_e)^T {\kappa}
 +{\kappa}(Y_e^\dagger Y_e)
 +6\,\Tr( Y_u^\dagger Y_u)\,{\kappa} 
 -2 g_1^2 {\kappa}- 6 g_2^2{\kappa}
 \; ,
\end{equation}
confirming the existing MSSM result \cite{Chankowski:1993tx,Babu:1993qv}.

\section{Conclusion}

A recent check of the SM \(\beta\)-function \cite{Chankowski:2001hx} showed
that previously published results were not quite correct.
Therefore, in this letter we have derived the RGE's for the effective 
dimension 5 operators which yield a 
Majorana mass for neutrinos after the electroweak symmetry breaking
in four types of 2HDM's and in the MSSM. For the MSSM, we confirmed 
the earlier result \cite{Chankowski:1993tx,Babu:1993qv}.
However, when we applied our general result \eqref{eq:2HDMBetaFunctions}
to the 2HDM discussed in \cite{Babu:1993qv}, we found that the non-diagonal
part of one of the \(\beta\)-functions, relevant for the evolution of the mixing
angles, is enhanced by a factor of 3. 

\filbreak
\ack

This work was supported by the 
``Sonderforschungsbereich~375 f\"ur Astro-Teilchenphysik der 
Deutschen Forschungsgemeinschaft''.
M.R. acknowledges support from the ``Promotionsstipendium des Freistaats Bayern''.

\endack

\end{document}